\documentclass[sigconf]{acmart}
\renewcommand\footnotetextcopyrightpermission[1]{} 

\usepackage{multirow}
\usepackage{algorithm}      
\usepackage{algpseudocode}  

\AtBeginDocument{%
  \providecommand\BibTeX{{%
    \normalfont B\kern-0.5em{\scshape i\kern-0.25em b}\kern-0.8em\TeX}}}

\begin{document}

\title{Machine Learning for Food Review and Recommendation}

\author{Le Tan Khang}
\affiliation{%
  \institution{Nanyang Technological University}
  \streetaddress{50 Nanyang Avenue}
  \country{Singapore}
  \postcode{639798}}
\email{tankhang001@e.ntu.edu.sg}

\author{Hui Siu Cheung}
\affiliation{%
  \institution{Nanyang Technological University}
  \streetaddress{50 Nanyang Avenue}
  \country{Singapore}
  \postcode{639798}}
\email{asschui@ntu.edu.sg}


\begin{abstract}
Food reviews and recommendations have always been important for online food service websites. However, reviewing and recommending food is not simple as it is likely to be overwhelmed by disparate contexts and meanings. In this paper, we use different deep learning approaches to address the problems of sentiment analysis, automatic review tag generation, and retrieval of food reviews. We propose to develop a web-based food review system at Nanyang Technological University (NTU) named NTU Food Hunter, which incorporates different deep learning approaches that help users with food selection. First, we implement the BERT and LSTM deep learning models into the system for sentiment analysis of food reviews. Then, we develop a Part-of-Speech (POS) algorithm to automatically identify and extract adjective-noun pairs from the review content for review tag generation based on POS tagging and dependency parsing. Finally, we also train a RankNet model for the \mbox{re-ranking} of the retrieval results to improve the accuracy in our Solr-based food reviews search system. The experimental results show that our proposed deep learning approaches are promising for the applications of real-world problems.
\end{abstract}




\maketitle
\pagestyle{empty} 

\section{Introduction}
Like product reviews in online shopping websites, food reviews are commonly available in most food review websites such as HungryGoWhere, Burpple, and Quandoo. They provide useful information for food lovers to know the opinions of others who have patronized the restaurants before about the restaurants and their food. As such, these websites help potential customers to decide whether they would like to eat at the restaurant. However, there are plenty of reviews on the website and it will cause information overloading for users to read all the reviews one by one. To tackle such a problem, it is important to mine the review data, and extract and provide useful information about the restaurants and food items for users. Recently, with the rapid development of machine learning and natural language processing techniques, these techniques have been applied to food review data for sentiment analysis and information retrieval. 

Therefore, in this paper, we apply machine learning techniques for analyzing food review data for sentiment analysis, automatic review tag generation, and reviews retrieval. In particular, we focus on analyzing the food review data from a website containing food reviews on the restaurants and food stalls in Nanyang Technological University (NTU) which has more than 20 canteens and restaurants. Overall, our work has achieved the followings: (1) We have developed the BERT and LSTM deep learning models for sentiment analysis of food reviews. (2) We have developed a Part-of-Speech (POS) algorithm for the extraction of adjective-noun pairs from review data to generate the review tags that label the reviews. (3) We have developed a ranking model based on the RankNet algorithm for the re-ranking of the retrieval results in the Solr-based food reviews search system which was developed for experimental purposes.

In this paper, we discuss the different machine learning techniques applied for food reviews in our NTU Food Hunter website. The rest of the paper is organized as follows. Section 2 reviews the related work. Section 3 discusses the proposed sentiment analysis techniques for NTU food reviews. Section 4 presents the proposed review tag generation approach. Section 5 discusses the proposed food review data retrieval approach. Finally, Section 6 concludes the paper.

\section{Related Works}
In this section, we review the related works on sentiment analysis, adjective-noun pair extraction, and information retrieval and Learning To Rank.

\subsection{Sentiment Analysis}
In recent years, deep learning models have played an increasingly important role in Natural Language Processing (NLP) as well as sentiment analysis. Since \citet{mikolov2010recurrent} proposed a statistical language model known as the Elman network \cite{elman1990finding} in 2010, Recurrent Neural Networks (RNNs) have gained much attention in NLP. However, the combination of previous hidden activations with the current inputs through addictive function in the Elman network is not powerful enough to deal with complex sentiment expressions \cite{wang2015predicting}. On the other hand, Long Short-Term Memory (LSTM) architecture \cite{hochreiter1997long} is shown to be an effective solution for tackling this problem. Additionally, multiplicative operations between word embeddings through gate structures provide more flexibility to the LSTM architecture. As a result, such composition outperforms the simple RNN architecture. For instance, the work of predicting sentiments in tweets by composing word embeddings with the LSTM recurrent neural network by
\citet{wang2015predicting} achieved a better performance result than the simple
RNN.

Furthermore, there have also been many breakthroughs in several NLP tasks by applying the method of pre-training language models on a large network with a large amount of unlabeled data and fine-tuning in downstream tasks. These works include OpenAI GPT \cite{radford2018improving} and BERT \cite{devlin2018bert}. Particularly, BERT is pre-trained on \textit{Masked Language Model Task} and \textit{Next Sentence Prediction Task} via a large cross-domain corpus. Unlike other representation models, the design of BERT is to pre-train deep bidirectional representations from unlabeled texts by jointly conditioning on both left and right context in all layers. As a result, \mbox{pre-trained} BERT is the first model that can be fine-tuned with just one additional output layer to obtain \mbox{state-of-the-art} results for a wide range of NLP tasks such as language inference and question answering \cite{devlin2018bert}. With the enormous potential of the fine-tuning method and BERT, \citet{sun2019fine} conducted further experiments on various fine-tuning methods of BERT on text classification and successfully achieved \mbox{state-of-the-art} performance on eight benchmark classification datasets including Yelp and IMDb.

\subsection{Adjective-Noun Pair Extraction}
An online review website usually offers a vast amount of useful information about a specific topic or a specific type of product. However, in many well-known review websites such as Yelp, the number of reviews for an individual item exceeds users’ ability to quickly skim through every line of reviews. Furthermore, these reviews are mostly unstructured and vary in length, writing style, and format. As a result, these problems prevent users from easily
grasping useful details about any item from online reviews. 

To tackle such problems, \citet{yatani2011review} proposed the Review Spotlight system. In this system, adjective-noun pairs were extracted from online reviews to summarize users’ opinions. With this feature, users do not have to read every review but only need to focus on these extracted word pairs. More specifically, adjective-noun pair extraction is the process of extracting all nouns in a document along with the adjectives modifying them. However, in the Review Spotlight implementation, this process was simply executed by labeling the part of speech for each word in a sentence and then pairing every noun with the closest adjective, which may be incorrect. For example, for the review "\emph{The food from this beautiful restaurant is awful.}", the "\emph{beautiful-food}" tag is incorrectly extracted instead of the "\emph{awful-food}" tag. Moreover, negation word detection is also not supported in the Review Spotlight system. For instance, the "\emph{not-good-food}" tag should be extracted from the review "\emph{This food is not good at all.}", instead of the \mbox{"\emph{good-food}"} tag. Therefore, in our proposed food review system, we further enhance the POS algorithm to correctly identify and extract adjective-noun word pairs as labeling tags of food review.

\subsection{Information Retrieval and Learning To Rank}
Information retrieval (IR) is the process of searching for documents of an unstructured nature, which is usually text, satisfying an information need from large collections of documents \cite{manning2008introduction}. In an IR system, when a user inputs a query, the system will return a ranked list of documents. The ranking of results is based on the relevance to the user's query. A good IR system should demonstrate a good ranking of retrieved documents. Fundamentally, \textit{relevance} is based on the textual similarity between an information requirement (\textit{query}) and an article (\textit{document}) \cite{cooper1971definition}. Specifically, every term in a query is assigned a weight and matched with document terms. The sum of weights represents the relevance score of a document to a query; the higher the score is, the more relevant a document is. BM25 \cite{robertson1995okapi} is one of the most widely used scoring functions which uses such term weighting method. To improve the accuracy of the ranking of the retrieved results, Learning To Rank (LTR) applies machine learning methods to train a ranking model to rank documents. Learning To Rank refers to the machine learning methods, either supervised, semi-supervised, or reinforcement learning, to construct ranking models for IR systems \cite{liu2011learning}. Then, the ranking model will re-rank the top $N$ retrieved documents. 

Successive studies and experiments on LTR have been conducted, many of which produced significant results. Generally, there are three main types of LTR approaches, namely the pointwise approach, the pairwise approach, and the listwise approach \cite{liu2011learning}. These algorithms can also be further categorized into their underlying learning techniques, such as the Support Vector Machine (SVM) techniques, the Boosting techniques, and the Neural Network techniques. Among the three approaches, the pairwise and listwise approaches usually can achieve better performance than the pointwise approach \cite{liu2011learning}. Ranking in IR is more about predicting relative order than accurate relevance degree. However, with the input of a single document, the pointwise approach is unable to consider the relative order between documents. In contrast, the pairwise approach has successfully addressed the ranking problems in many IR systems. For instance, \citet{burges2005learning} used RankNet for large-scale web search, while \citet{cao2006adapting} applied Ranking SVM to document retrieval by using a modified loss function.

\section{Sentiment Analysis}
In this section, we describe the implementation of our LSTM and BERT neural network models for sentiment analysis of food reviews.

\subsection{Proposed LSTM Model}
\begin{figure}[h]
    \centering
    \includegraphics[width=\linewidth]{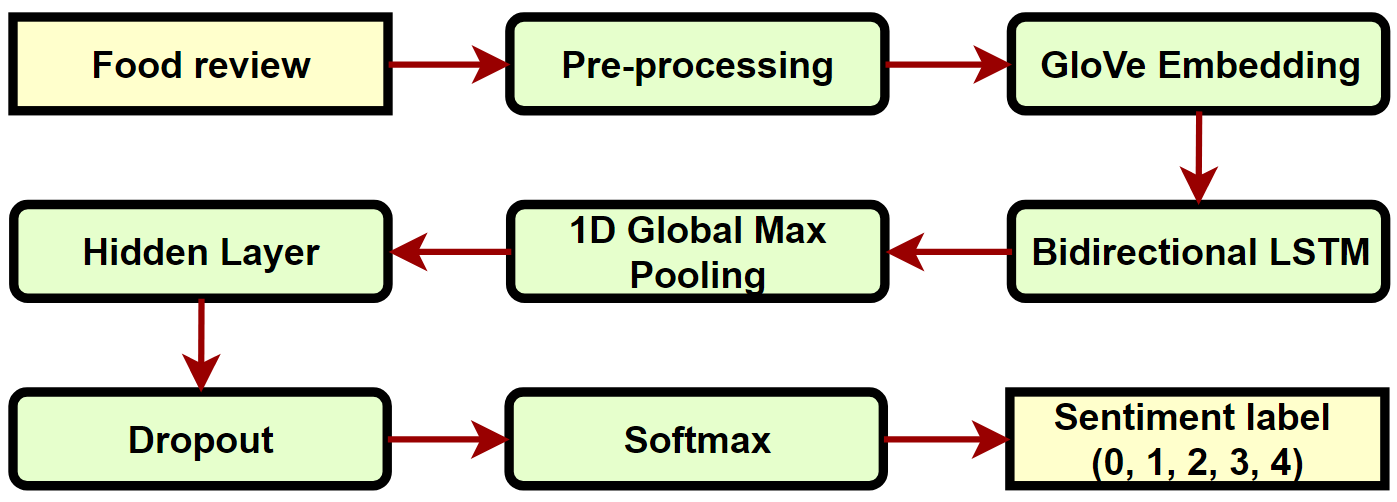}
    \caption{Proposed LSTM architecture.}
    \label{fig: proposed-lstm-architecture}
    \Description{The architecture of the proposed LSTM model, including key components such as GloVe Embedding and Bidirectional LSTM.}
\end{figure}

In this paper, we have proposed the LSTM RNNs for sentiment analysis of food reviews as they generally can achieve better performance than the traditional RNNs for learning relationships in sequential data. Figure~\ref{fig: proposed-lstm-architecture} shows the proposed LSTM model architecture.

In our LSTM architecture, after the pre-processing step, a food review is transformed to a token sequence of length 500 for the model input. Subsequently, we prepare an embedding matrix such that for each token at index \textit{i}, there is a corresponding GloVe embedding vector contained at index \textit{i} of the matrix. The embedding matrix is then loaded into the embedding layer of size 50 and fed through our LSTM model. Furthermore, in the design, we use a bidirectional LSTM layer of size 100, followed by a one-dimensional global max pooling layer and a hidden layer of size 50 with the ReLU (Rectified Linear Unit) activation function. To prevent overfitting in the model training process, a dropout value of 0.1 is selected to randomly turning off 10\% of the nodes in the networks. The Adam optimizer is chosen as it is one of the most popular gradient descent optimization algorithms for training deep learning models. The output layer of the model is the Softmax function for the multi-class classification of reviews. Table~\ref{table:lstm-hyperparameters} shows the settings of the hyperparameters of our proposed LSTM model. 

\begin{table}
    \caption{Hyperparameters of the LSTM model.}
    \label{table:lstm-hyperparameters}
    \begin{center}
        \begin{tabular}{ l l } 
            \toprule
            \multicolumn{1}{c}{\textbf{Description}} & \multicolumn{1}{c}{\textbf{Value}} \\
            \midrule
            Input length & 500 \\ 
            Embedding size & 50 \\
            Word embeddings & GloVe 50d word vectors \\
            Bidirectional LSTM size & 100 \\
            Pooling strategy & 1D global max pooling \\
            Hidden layer size & 50 \\
            Activation function & Rectified Linear Unit \\
            Dropout rate & 0.1 \\
            Recurrent Dropout rate & 0.1 \\
            Optimizer & Adam \\
            Output & Softmax \\
            \bottomrule
        \end{tabular}
    \end{center}
\end{table}

\subsection{Proposed BERT Model}
\begin{figure}[h]
    \centering
    \includegraphics[width=\linewidth]{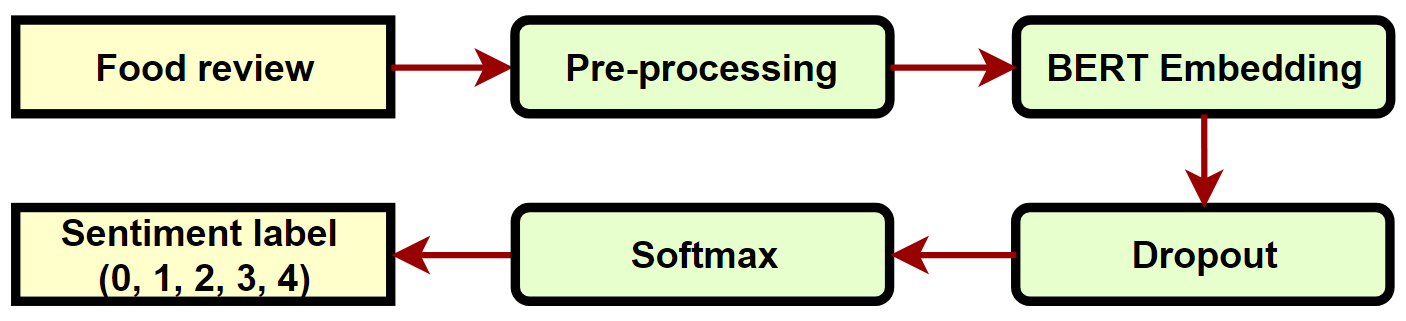}
    \caption{Proposed BERT architecture.}
    \label{fig: proposed-bert-architecture}
    \Description{The architecture of the proposed BERT model, including key components such as BERT Embedding.}
\end{figure}

The BERT model, proposed by \citet{devlin2018bert}, adopts many innovations in the NLP field, namely contextualized word representations \cite{peters2018deep}, the transformer architecture \cite{vaswani2017attention}, and pre-training on a language modeling task with subsequent end-to-end fine-tuning on a downstream task \cite{radford2018improving,howard2018universal}. There are two parameter intensity settings for BERT: \textbf{BERT$_\text{BASE}$} and \textbf{BERT$_\text{LARGE}$}. In this paper, due to our hardware limitation, we use the BERT$_\text{BASE}$ model for the sentiment classification of food reviews. As the pre-trained BERT model can be fine-tuned easily, we build a simple architecture to show that BERT can achieve promising results without complicated architectures for any specific task. Figure~\ref{fig: proposed-bert-architecture} shows the proposed BERT model architecture.

In our BERT architecture, after the tokenization process, a food review is transformed to a token sequence of length 512 for the model input. In our design, we use a BERT embedding layer of size 256 with the ReLU (Rectified Linear Unit) activation function. Subsequently, a dropout layer is added with a value of 0.1 to prevent the model from overfitting. Being the most popular gradient descent optimization algorithm for training deep learning models, the Adam optimizer is utilized in our model. Lastly, for the multi-class classification of reviews, we adopt the Softmax function for the output layer. Table~\ref{table:bert-hyperparameters} shows the settings of the hyperparameters of our proposed fine-tuned BERT model.

\begin{table}
    \caption{Hyperparameters of the BERT model.}
    \label{table:bert-hyperparameters}
    \begin{center}
        \begin{tabular}{ l l } 
            \toprule
            \multicolumn{1}{c}{\textbf{Description}} & \multicolumn{1}{c}{\textbf{Value}} \\
            \midrule
            Input length & 512 \\ 
            BERT embedding layer size & 256 \\
            Activation function & Rectified Linear Unit \\
            Dropout rate & 0.1 \\
            Optimizer & Adam \\
            Output & Softmax \\
            \bottomrule
        \end{tabular}
    \end{center}
\end{table}

\subsection{Performance Evaluation}
For the experiments on sentiment analysis, we use the Yelp reviews dataset, which is a widely used benchmark in the deep learning community \cite{zhang2015character}. In addition, we consider only the reviews that are categorized as food or restaurants in the dataset. Table~\ref{table:dataset-statistics} shows the statistics of both the training and test sets from the Yelp dataset.

\begin{table}
    \caption{Dataset statistics of the training and test sets.}
    \label{table:dataset-statistics}
    \begin{center}
        \begin{tabular}{ l l l } 
            \toprule
            \multicolumn{1}{c}{\textbf{Label}} & \multicolumn{1}{c}{\textbf{Training set}} &
            \multicolumn{1}{c}{\textbf{Test set}} \\
            \midrule
            0 & 59057 & 4270 \\
            1 & 68007 & 4961 \\
            2 & 108800 & 7712 \\
            3 & 228393 & 16203 \\
            4 & 235743 & 16854 \\
            \midrule
            \textbf{Total} & 700000 & 50000 \\
            \bottomrule
        \end{tabular}
    \end{center}
\end{table}

Table~\ref{table:compare-lstm-bert} shows the performance results of our proposed LSTM and BERT models for both training and testing. Our proposed LSTM model is trained for 4 epochs and achieves an accuracy of around 66.76\%. On the other hand, our proposed BERT model only needs to be trained for 2 epochs and is able to achieve a higher accuracy of 70.52\%. Overall, our proposed fine-tuned BERT model is able to achieve promising results by taking the advantage of the pre-trained BERT model with many innovative features in the NLP field.

\begin{table}
    \caption{Performance results on sentiment analysis.}
    \label{table:compare-lstm-bert}
    \begin{center}
        \begin{tabular}{ l l l l l } 
            \toprule
            \multirow{2}{*}[-2pt]{\textbf{Model}} & \multicolumn{2}{c}{\textbf{train}} & \multicolumn{2}{c}{\textbf{test}} \\
                                            \cmidrule(lr){2-3} \cmidrule(lr){4-5}
                                            & \multicolumn{1}{c}{accuracy} & \multicolumn{1}{c}{loss} & \multicolumn{1}{c}{accuracy} & \multicolumn{1}{c}{loss} \\
            \midrule
            \textbf{LSTM} & 0.6794 & 0.7251 & 0.6676 & 0.7520 \\
            \textbf{BERT} & 0.7189 & 0.6338 & 0.7052 & 0.6656 \\ 
            \bottomrule
        \end{tabular}
    \end{center}
\end{table}

\section{Review Tag Generation}
This section proposes an approach to generate the review tags for each review comment. This serves as a tag to represent the corresponding review. Each review tag will carry the sentiment according to the review comment. The proposed approach for automatic review tag generation is based on adjective-noun pair extraction. In this section, we present the algorithm for adjective-noun pair extraction.

\subsection{Proposed Algorithm}
Our proposed algorithm for extracting adjective-noun pairs is built based on the principles of POS tagging and dependency parsing. Dependency parsing is the task of extracting a dependency parse of a sentence that represents its grammatical
structure. It defines the relationship between the "head" words and words, which modifies those heads \cite{kubler2009dependency}. 

SpaCy is a free open-source library for advanced NLP in Python \cite{spacy}. It provides support for various applications such as information extraction, natural language understanding, and text pre-processing for deep learning. There are many features in spaCy such as tokenization, POS tagging, dependency parsing, named entity recognition, etc. In our proposed algorithm, we use the POS tagging and dependency parsing features from spaCy to analyze the grammatical structure of each review.

\begin{figure}[h]
    \centering
    \includegraphics[width=\linewidth]{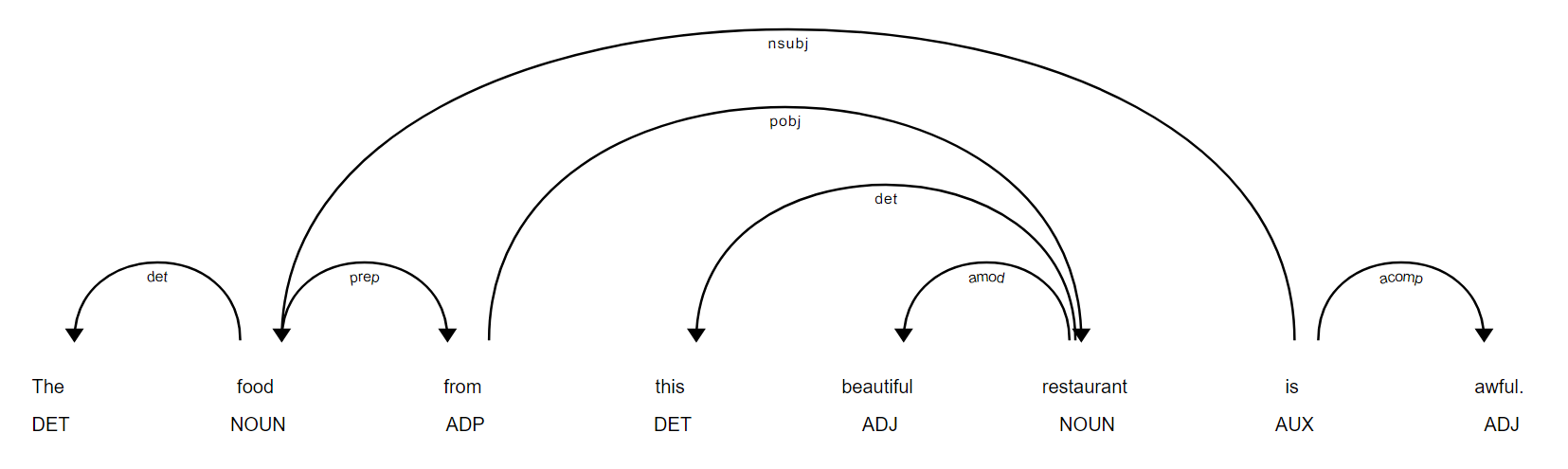}
    \caption{Dependency parsing of a sample food review.}
    \label{fig: sample-dep-parse}
    \Description{The POS tags and dependency relationships between words in a food review.}
\end{figure}

After performing the POS tagging and dependency parsing processes, our algorithm traverses the parsed text with its labeled POS tags and dependencies to identify and extract adjective-noun word pairs. Specifically, the algorithm searches for all adjectives and then checks each adjective whether its dependency is "amod" (adjectival modifier) or "acomp" (adjectival complement). If the dependency is "amod," the adjective and its "head" word are returned as a pair. Otherwise, for the "acomp" dependency, the subtree of the "head" word of the adjective is generated. The algorithm searches the subtree for a noun with the "nsubj" (nominal subject) dependency and returns it along with the adjective as a pair. Last but not least, in the subtree of the "head" word of the adjective, if a negation word is found, it will be added to the front of the current adjective-noun pair. Finally, the algorithm will output a list of adjective-noun pairs extracted from the input text. For example, Figure~\ref{fig: sample-dep-parse} shows the extraction process on a sample food review "\emph{The food from this beautiful restaurant is awful.}" Particularly, our algorithm detects two adjectives "\emph{beautiful}" and "\emph{awful}" as well as their "amod" and "acomp" dependencies with "\emph{restaurant}" and "\emph{food}" respectively. As a result, the two extracted adjective-noun pairs are \mbox{"\emph{beautiful-restaurant}"} and \mbox{"\emph{awful-food}."} 

\begin{figure}[h]
  \includegraphics[width=\linewidth]{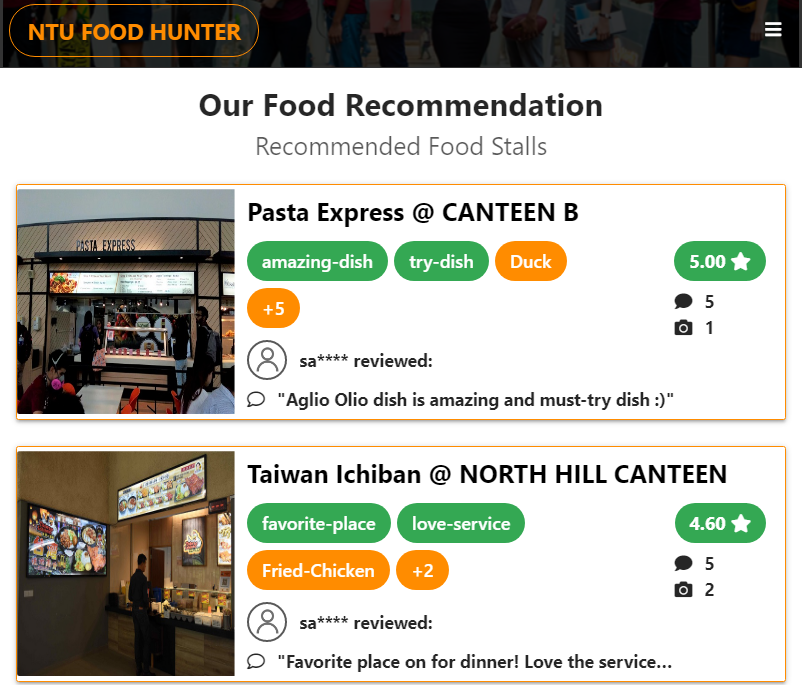}
  \caption{Review tags in the NTU Food Hunter website.}
  \label{fig: ntu-food-hunter-snapshot}
  \Description{The home page of our food review website, named NTU Food Hunter. The information of two food services are displayed.}
\end{figure}

In our NTU food review website, adjective-noun pairs extracted from food reviews are displayed in the form of tags to help users to have a quick view of opinions about any food services. Additionally, these review tags also indicate the sentiments of the reviews, in which negative, neutral, and positive sentiments are color-coded as red, yellow and green, respectively. Figure~\ref{fig: ntu-food-hunter-snapshot} shows the home page of our NTU Food Hunter website, including several review tags.

\subsection{Performance Evaluation}
To evaluate the performance of the proposed adjective-noun pair extraction algorithm for review tag generation, we use 539 food reviews that are obtained from the Yelp reviews dataset for the evaluation. In the experiment, we run the proposed algorithm for the 539 food reviews and generate the adjective-noun pairs. We then inspect the extraction results manually by a group of evaluators who are experienced in the NLP field or fluent in English. As a result, we have achieved the performance result of 95\%. 

In the experiment, we have also observed that the proposed algorithm is unable to extract the adjective-noun pairs correctly for certain review data. This is mainly due to the reason that some of the food reviews are incoherent or grammatically incorrect, and therefore SpaCy is unable to label the corresponding text with correct POS tags and dependencies. Overall, our proposed algorithm is a promising approach for online review tag generation from food reviews.

\section{Food Reviews Search}
In this section, we discuss our proposed food reviews search system which is based on a LTR model, using the RankNet algorithm, for the re-ranking of Solr-based retrieval results.

\subsection{Retrieval and Ranking}
The retrieval and ranking of food reviews are based on the similarity between the search query and food reviews. As such, we need to identify features and define the similarity scoring function for each feature. In this paper, we define the following three types of similarity features in the food reviews search system.

\subsubsection{Textual Similarity} One of the most fundamental approaches to decide whether two pieces of text are identical is to calculate their textual similarity. As in many basic information retrieval systems, a ranking function, such as tf-idf or BM25, is used to measure the relevance of documents to a given search query based on the appearance of query terms in each document. For our model, we use BM25 (based on Solr \cite{solr}) to score textual similarity between a query and a review.

\subsubsection{Semantic Similarity} A search system needs to measure the relevance of a document and a query beyond the simple textual similarity. Specifically, textual similarity at the word level does not take into consideration the actual meaning of words or the entire phrase in context. As such, we shall model the semantic similarity between two pieces of text to achieve a better search result. The traditional approach to address semantic search is to transform each sentence into a vector space such that semantically similar sentences will be close to each other. We use Sentence-BERT (SBERT) to derive semantically meaningful BERT embeddings that can be compared using
cosine similarity \cite{reimers2019sentence}.

\subsubsection{Food Category Similarity} With both textual and semantic similarities, our food reviews search system can retrieve the most relevant documents in most cases. However, problems arise if users want to search for reviews from a particular food category or item. To resolve this issue, for each review, we create a list of related food categories as a feature, and then use BM25 to compute the similarity score.

\subsection{Learning To Rank for Re-ranking}
\begin{figure}[h]
    \centering
    \includegraphics[width=\linewidth]{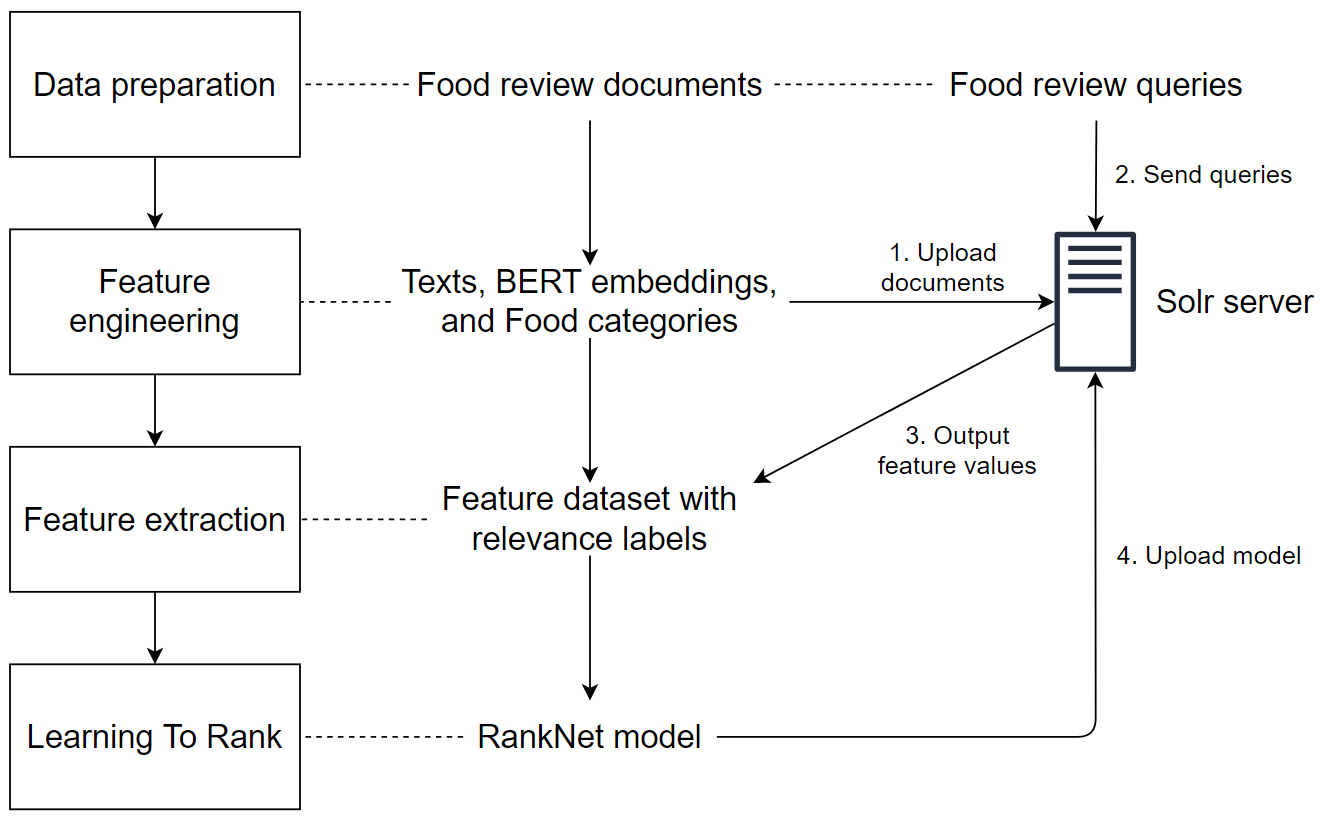}
    \caption{Development process of the ranking model.}
    \label{fig: dev-process-overview}
    \Description{Four main steps in the development process of the ranking model with relevant details.}
\end{figure}

To improve the food reviews search system, we have incorporated a LTR model based on RankNet. Figure~\ref{fig: dev-process-overview} shows the development process of the proposed model for our food reviews search system, which consists of the following steps:

\begin{enumerate}
    \item Data preparation: From the food reviews data, we create a set of queries and documents, in which each document represents a review.
    \item Feature engineering: For each document, we define its features and their similarity scoring function. As such, given a query and a document, their feature similarity score, also known as feature value, is computed to measure the query-document relevance. We then upload the documents, including their defined features, to Solr for the next step of extracting feature values.
    \item Feature extraction: We send the created queries to Solr. From these queries and the uploaded documents, Solr extracts and returns feature values for query-document pairs. Along with extracted feature values, we determine the relevance label, either 0 or 1, for each query-document pair to build a feature dataset for the training and testing of our ranking model.
    \item Learning To Rank: We train the RankNet model using the complete feature dataset, consisting of feature values and relevance labels. The model will learn to perform the re-ranking of retrieval results based on query-document feature values such that relevant query-document pairs are ranked higher than irrelevant ones.
\end{enumerate}

\begin{figure}[h]
    \centering
    \includegraphics[width=\linewidth]{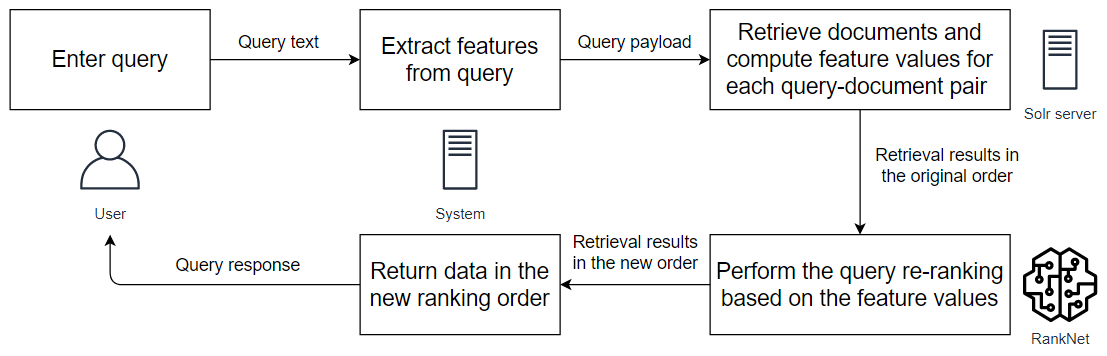}
    \caption{Search query process using the RankNet model.}
    \label{fig: ltr-query-process}
    \Description{End-to-end process of a Learning To Rank search query, from user's input to ranked retrieval results.}
\end{figure}

The retrieval process is then carried out as follows. First, when a user enters a query, the search system will extract features from it, such as the review text and its BERT embedding. Then, the query, along with all its features, is sent to the Solr server using the payload format. Solr first retrieves documents and ranks them by their cosine similarity to the BERT embedding of the query. Next, Solr will match query features with the corresponding fields in each document to compute all feature values. The retrieval results in the original order are then fed into the RankNet model to perform the query re-ranking based on their computed feature values. Subsequently, the RankNet model will output the retrieval results in a new order. Finally, the search system returns these retrieved data, in the new ranking order, to the user as the query response. Figure~\ref{fig: ltr-query-process} illustrates the process of executing search query using the RankNet model.

\subsection{Performance Evaluation}
We conduct an experiment and evaluate the performance of our proposed search system by comparing it with the classical retrieval methods such as tf-idf and BM25. Specifically, for each retrieval method, we execute 10 test queries to retrieve more than 300 documents from Solr. Subsequently, by using the predefined criteria, the relevance label of each \mbox{query-document} pair is determined. Finally, as the relevance labels in our data are binary (either 1 or 0), we use the Mean Average Precision at Position \textit{k} (MAP@\textit{k}) and Mean Reciprocal Rank (MRR) metrics to evaluate the retrieval results of each method.

Table~\ref{table:ir-performance-comparison} shows that our search system has performed significantly better than other methods. For all metrics, our proposed search system has improved the performance by between 29\% and 46\% when compared to BM25 and tf-idf respectively. One intuitive explanation is that the purpose of the food reviews search system is to find reviews with the most similar meaning, simple text retrieval-based methods are not sufficient as they only consider the textual information. In contrast, our RankNet model evaluates both textual and semantic aspects of a review as well as the food category information to find the most similar reviews. This shows that our proposed search system is effective for the retrieval of food review data.

\begin{table}
    \caption{Performance results on food reviews search.}
    \label{table:ir-performance-comparison}
    \begin{center}
        \begin{tabular}{ p{1.5cm} p{1.2cm} p{1.2cm} p{1.2cm} p{1.2cm} } 
            \toprule
            & \multicolumn{1}{c}{\textbf{MAP@1}} & \multicolumn{1}{c}{\textbf{MAP@3}} &
            \multicolumn{1}{c}{\textbf{MAP@5}} &
            \multicolumn{1}{c}{\textbf{MRR}} \\
            \midrule
            tf-idf & 0.6000 & 0.4722 & 0.4713 & 0.7021 \\
            BM25 & 0.6000 & 0.6389 & 0.5793 & 0.6795 \\
            \textbf{Ours} & 1.0000 & 0.9333 & 0.9220 & 1.0000 \\
            \bottomrule
        \end{tabular}
    \end{center}
\end{table}

\section{Conclusion}
In this paper, we have applied different deep learning techniques for food reviews and recommendations in our proposed food review website, named NTU Food Hunter. First, for the sentiment analysis of food reviews, we incorporated our trained LSTM and BERT deep learning models into the system. Second, we implemented a POS algorithm based on \mbox{adjective-noun} pair extraction for automatic review tag generation. Last but not least, in our Solr-based food reviews search system, we trained and integrated a RankNet model to improve the accuracy of the ranking of the retrieval results. The experimental results have shown that our proposed deep learning approaches are useful to address real-world problems.

\bibliographystyle{ACM-Reference-Format}
\bibliography{sample-base}

\end{document}